\documentclass[conference]{IEEEtran}

\usepackage{graphicx}
\usepackage{fixltx2e}
\usepackage{float}

\IEEEoverridecommandlockouts
\usepackage{cite}
\usepackage{amsmath,amssymb,amsfonts}
\usepackage{algorithmic}
\usepackage{textcomp}
\usepackage{xcolor}
\def\BibTeX{{\rm B\kern-.05em{\sc i\kern-.025em b}\kern-.08em
    T\kern-.1667em\lower.7ex\hbox{E}\kern-.125emX}}

\usepackage{courier}

\usepackage{listings,xcolor}
\usepackage{url}

\begin{document}

\title{Processing Columnar Collider Data with GPU-Accelerated Kernels
\thanks{This work is supported by the the Prime NSF Tier2 award 1624356 and the Caltech HEP DOE award DE-SC0011925.}
}
%\author{
%    \IEEEauthorblockN{Joosep~Pata, Maria~Spiropulu\IEEEauthorrefmark{1}}\\
%    \IEEEauthorblockA{\IEEEauthorrefmark{1}California Institute of Technology
%    \\\{jpata, smaria\}@caltech.edu}
%}

% The paper headers
%\markboth{}%
%{HEPAccelerate: Fast Analysis of Columnar Collider Data}

\author{\IEEEauthorblockN{Joosep Pata}
\IEEEauthorblockA{\textit{Department of Physics, Mathematics and Astronomy} \\
\textit{California Institute of Technology}\\
Pasadena, California, USA \\
jpata@caltech.edu}
\and
\IEEEauthorblockN{Maria Spiropulu}
\IEEEauthorblockA{\textit{Department of Physics, Mathematics and Astronomy} \\
\textit{California Institute of Technology}\\
Pasadena, California, USA \\
smaria@caltech.edu}
}

\maketitle

\begin{abstract}
At high energy physics experiments, processing billions of records of structured numerical data from collider events to a few statistical summaries is a common task. The data processing is typically more complex than standard query languages allow, such that custom numerical codes are used. At present, these codes mostly operate on individual event records and are parallelized in multi-step data reduction workflows using batch jobs across CPU farms. Based on a simplified top quark pair analysis with CMS Open Data, we demonstrate that it is possible to carry out significant parts of a collider analysis at a rate of around a million events per second on a single multicore server with optional GPU acceleration. This is achieved by representing HEP event data as memory-mappable sparse arrays of columns, and by expressing common analysis operations as kernels that can be used to process the event data in parallel. We find that only a small number of relatively simple functional kernels are needed for a generic HEP analysis. The approach based on columnar processing of data could speed up and simplify the cycle for delivering physics results at HEP experiments. We release the \texttt{hepaccelerate} prototype library as a demonstrator of such methods.
\end{abstract}

\maketitle

\section{Introduction}
At the general-purpose experiments of the Large Hadron Collider such as CMS or ATLAS, the final stage of data processing is typically carried out over several terabytes of numerical data residing on a shared cluster of servers via batch processing.
The data consist of columns of physics related variables or features for the recorded particles such as electrons, muons, jets and photons for each event in billions of rows.
In addition to the columns of purely kinematic information of particle momentum, typically expressed in spherical coordinates of $p_T$, $\eta$, $\phi$ and $M$, each particle carries a number of features that describe the reconstruction details and other high-level properties of the reconstructed particles.
For example, for muons, we might record the number of tracker layers where an associated hit was found, whether or not it reached the outer muon chambers and in simulation the index of the associated generator-level particle, thereby cross-linking two collections.
A typical event might contain hundreds of measured or simulated particles, such that compressed event sizes for such reduced data formats are on the order of a few kilobytes per event.
In practice, mixed precision floating points are used to store data only up to experimental precision, such that the number of features per event is on the order of a few hundred.

\begin{figure}[!tp]
  \includegraphics[width=1.0\linewidth]{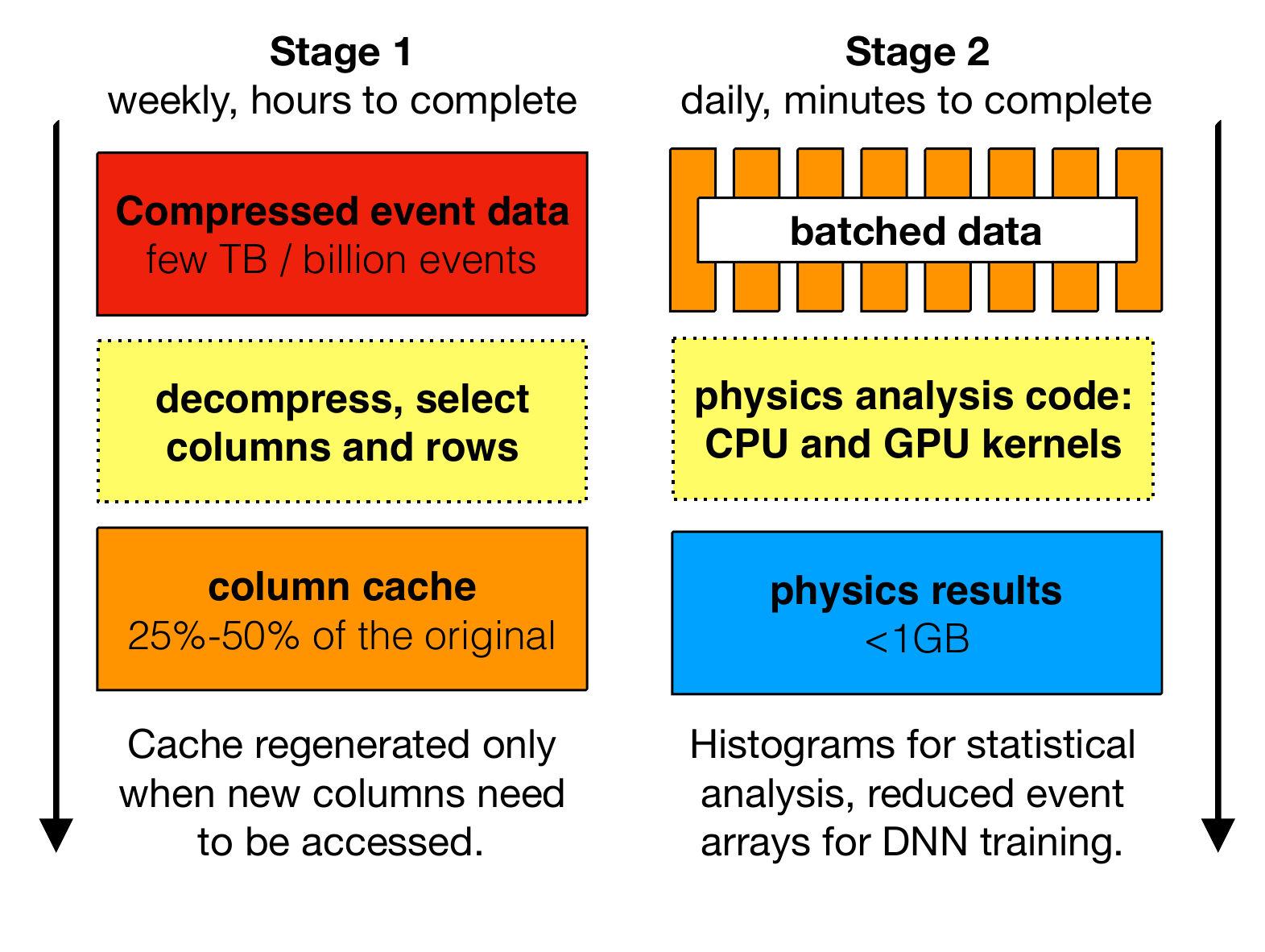}
  \caption{The flowchart of the accelerated workflow for an example analysis. In the decompression stage, done on a weekly basis as new data samples arrive, the necessary columns from the compressed ROOT files of a few TB are saved to a smaller uncompressed cache to minimize decompression overhead at successive analysis iterations. The size of this uncompressed cache varies, but is generally around 20-50\% of the original compressed data. In the analysis stage, which is done on an hourly basis, the cache is loaded in batches of a few GB of contiguous 1-dimensional arrays corresponding to sparse event content, with kernels being dispatched on the data either in host or device memory. The final result is a significantly reduced statistical summary of the original data, typically in the form of hundreds of histograms.}
  \label{fig:flowchart}
\end{figure}

A typical physics analysis at the LHC such as the precision measurement of a particle property or the search for a new physics process involves billions of recorded and simulated events.
The final result typically requires tens to hundreds of iterations over this dataset while the analysis is ongoing over a few months to a year. For each iteration of the analysis, hundreds of batch jobs of custom reduction software is run over these data.
By reducing the complexity and increasing the speed of these workflows, HEP experiments could deliver results from large datasets with faster turn-around times and correspondingly more time spent in development of new analysis methods as well as validation and verification, accelerating discovery and making the results more robust.

Recently, heterogeneous and highly parallel computing architectures beyond consumer x86 processors such as GPUs, TPUs and FPGAs have become increasingly prevalent in scientific computing. We investigate the use of array-based computational kernels that are well-suited for parallel architectures for carrying out the final data analysis in HEP. We have implemented a prototypical top quark pair analysis involving the selection and calibration of jets in an event, along with the evaluation of a machine learning discriminator. Although we use a specific and relatively simple analysis as an example, the approach based on array computing with accelerated kernels is generic and can be used for other collider analyses. The purpose of this report is to document the efforts of processing terabyte-scale data in HEP fast and efficiently. We release the \texttt{hepaccelerate} library for further discussion~\cite{hepaccelerate}.

In the following, we explore the approach based on columnar data analysis using vectorizable kernels in more detail.
In section~\ref{sec:data}, we describe the structure of the data and discuss how data sparsity is handled efficiently.
We introduce physics-specific computational kernels in section~\ref{sec:kernels} and describe the measured performance under a variety of conditions in section~\ref{sec:performance}.
Finally, we conclude with a summary and outlook in section~\ref{sec:summary}.

\section{Data structure}
\label{sec:data}
We can represent HEP collider data in the form of two-dimensional matrices, where the rows correspond to events and columns correspond to features in the event such as the momentum components of all measured particles.
Due to the random nature of the underlying physics processes that produce a varying number of particles per event, the number of features varies from one collider event to another, such that a fixed-size two dimensional representation is not memory-efficient.
Therefore, the standard HEP software framework based on ROOT includes mechanisms for representing dynamically-sized arrays as well as complete C++ classes with arbitrary structure as the feature columns, along with a mechanism for serializing and deserializing these dynamic arrays~\cite{Antcheva:2009zz}.

\begin{figure}[!tp]
  \includegraphics[width=1.0\linewidth]{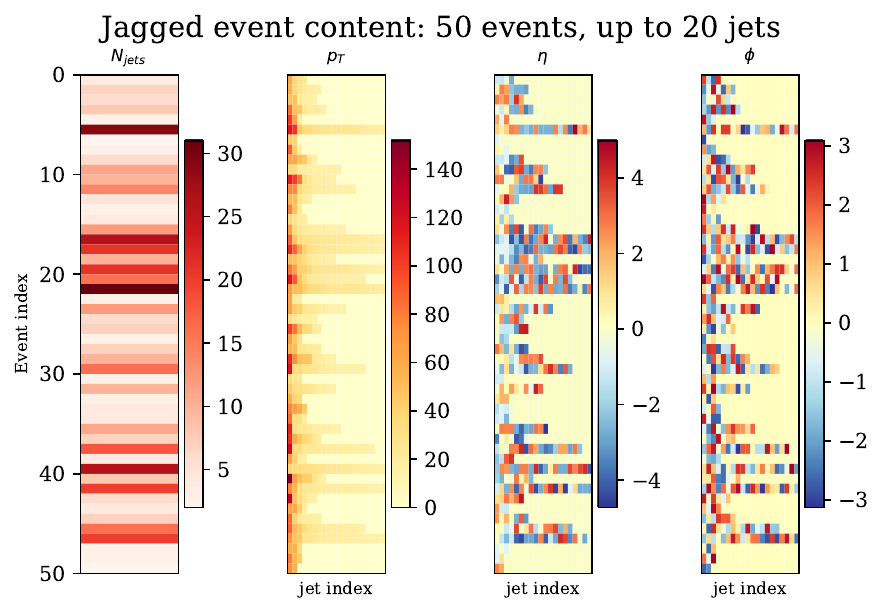}
  \caption{A visual representation of the jagged data structure of the jet $p_T$, $\eta$ and $\phi$ content in 50 simulated events.  On the leftmost figure, we show the number of jets per event, one event per row, derived from the offset array. In the three rightmost columns, we show the jet content in events, visualizing the $p_T$, $\eta$ and $\phi$ of the first 20 jets for each event.}
  \label{fig:jagged}
\end{figure}

Based on the approach first introduced in the \texttt{uproot}~\cite{uproot} and \texttt{awkward-array}~\cite{awkward} python libraries, many existing HEP data files with a varying number of particles per event can be represented and efficiently loaded as sparse arrays with an underlying one-dimensional array for a single feature.
Event boundaries are encoded in an offset array that records the particle count per event.
Therefore, the full structure of $N$ events, $M$ particles in total, can be represented by a contiguous offset array of size $N+1$ and a contiguous data array of size $M$ for each particle feature.
This can easily be extended to event formats where the event contains particle collections of different types, e.g. jets, electrons and muons.
By using the data and offset arrays as the basis for computation, efficient computational kernels can be implemented and evaluated on the data.
We illustrate this sparse or jagged data structure on Fig.~\ref{fig:jagged}.

In practice, analysis-level HEP event data are stored in compressed files of raw columnar features in a so-called ``flat analysis ntuple'' form, implemented via the ROOT library.
Out of hundreds of stored features, a typical analysis might use approximately 50 to 100, discarding the rest and only using them rarely for certain auxiliary calibration purposes.
When the same features are accessed multiple times, the cost of decompressing the data can be significant.
Therefore, in order to maximize the computational efficiency of the analysis, we have implemented a simple cache based on memory mapped files for the feature and offset data.
The efficiency of the uncompressed cache depends on the ratio of features used for analysis.
We find that for the benchmark analysis, the average uncompressed cache size is approximately 0.5 kB/event after choosing only the necessary columns, down from approximately 2 kB/event in compressed but unreduced form. 
The choice of compression algorithms can and should be addressed further in optimizing the file formats for cold storage and analysis~\cite{rootio}.

\section{Computational kernels}
\label{sec:kernels}
In the context of this report, a kernel is a function that is evaluated on the elements of an array to transform the underlying data.
A simple kernel could compute the square root of all the values in the array. More complicated kernels such as convolutions might involve relations between neighbouring array elements or involve multiple inputs and outputs of varying sizes.
When the individual kernel calls across the data are independent of each other, these kernels can be evaluated in parallel over the data using single-instruction, multiple-data (SIMD) processors.
The use of efficient computational kernels with an easy-to-use API has proven to be successful for machine learning frameworks such as \texttt{tensorflow}.
Therefore, we propose to implement HEP analyses similarly using common kernels that are steered from a single high-level code.
We note that the columnar data analysis approach based on single-threaded kernels is already recognized in HEP~\cite{coffea}.
Our contribution is to further extend the computational efficiency and scalability of the kernels to parallel hardware such as multi-threaded CPUs and propose a GPU implementation.

In the following, we will demonstrate that by using the data and offset arrays as described in section~\ref{sec:data}, common HEP operations can be formulated as kernels and dispatched to the computational hardware, thereby processing the event data in parallel.
The central result of this report is that only a small number of simple kernels, easily implemented in e.g. Python or C, are needed to implement a realistic HEP analysis.

A prototypical HEP-specific kernel would be to find the scalar sum $H_T=\sum_{i\in\mathrm{event}}p_T^i$ of all particles passing some quality criteria in an event.
We show the Python implementation for this on Listing~\ref{lst:sumht}.
This kernel takes as input the $M$-element data array of all particle transverse momenta \texttt{pt\_data} and an $N+1$-element array of the event offsets.
In addition, as we wish to include only selected particles in selected events in this analysis, we use an $N$-element boolean mask for events and $M$-element boolean mask for the particles that have passed selection.
These masks can be propagated to further functions, making it possible to efficiently chain computations without resorting to data copying.
Finally, the output is stored in a pre-allocated array of size $N$.

\begin{figure}[!t]
\begin{minipage}{0.95\linewidth}
\begin{center}
\vspace{0.1cm}
\begin{lstlisting}[frame=single,language=Python,caption={Python code for the kernel computing the scalar sum of selected particle momenta $H_T$. The inputs are \texttt{pt\_data}, an $M$-element array of $p_T$ data for all the particles, the $N+1$-element \texttt{offsets} array with the indices between the events in the particle collections, as well masks for events and particles that should be considered. On line 10, the kernel is executed in parallel over the events using the Numba \texttt{prange} iterator, which creates multithreaded code across the loop iterations. On line 19, the particles in the event are iterated over sequentially.},label={lst:sumht},abovecaptionskip=10pt]
def sum_ht(
  pt_data, offsets,
  mask_rows, mask_content,
  out):

  N = len(offsets) - 1
  M = len(pt_data)

  #loop over events in parallel
  for iev in prange(N):
    if not mask_rows[iev]:
      continue

    #indices of the particles in this event
    i0 = offsets[iev]
    i1 = offsets[iev + 1]

    #loop over particles in this event
    for ielem in range(i0, i1):
      if mask_content[ielem]:
        out[iev] += pt_data[ielem]
\end{lstlisting}
\end{center}
\end{minipage}
\end{figure}

This kernel generically reduces a data array using a pairwise operator $(+)$ within the offsets and can be reused in other cases, such as counting the number of particles per event passing a certain selection.
Other kernels that turn out to be useful are related to finding the minimum or maximum value within the offsets or retrieving or setting the $m$-th value of an array within the event offsets.

The generic kernels we have implemented for the benchmark top quark pair analysis are the following:
\begin{itemize}
  \item \verb|get_in_offsets|: given jagged data with offsets, retrieves the $n$-th elements for all rows. This can be used to create a contiguous array of e.g. the leading jet $p_T$ for further numerical analysis.  
  \item \verb|set_in_offsets|: as above, but sets the $n$-th element to a value. This can be used to selectively mask objects in a collection, e.g. to select the first two jets ordered by $p_T$.
  \item \verb|sum_in_offsets|: given jagged data with offsets, calculates the sum of the values within the rows. As we have described above, this can be used to compute a total within events, either to count objects passing selection criteria by summing masks or to compute observables such as $H_T$.
  \item \verb|max_in_offsets|: as above, but calculates the maximum of the values within rows.
  \item \verb|min_in_offsets|: as above, but calculates the minimum.
  \item \verb|fill_histogram|: given a data array, a weight array, histogram bin edges and contents, fills the weighted data to the histogram. This is used to create 1-dimensional histograms that are common in HEP. Extension to multidimensional histograms is straightforward.
  \item \verb|get_bin_contents|: given a data array and a lookup histogram, retrieves the bin contents corresponding to each data array element. This is used for implementing histogram-based reweighting.
\end{itemize}

This demonstrates that a small number of dedicated kernels can be used to offload a significant part of the analysis.
There are also a number of less generic analysis operations which do not easily decompose into other fundamental array operations, but are still useful for HEP analysis.
A particular example would be to find the first two muons of opposite charge in the event, or to perform $\Delta R(\eta, \phi)$-matching between two collections of particles.
In the standard approach, the physicist might simply write down procedural code in the form of a nested loop over the particles in the event which is terminated when the matching criterion is satisfied.
These functions can similarly be expressed in the form of a dedicated kernels that do a single pass over the data and that are easy to write down as well as debug procedurally for the physicist before being dispatched to parallel processors.
We choose to implement these explicitly rather than rely on more general array operations that perform cross-collection joins for execution speed and simplicity of implementation. These specialized kernels are as follows:

\begin{itemize}
  \item \verb|mask_deltar_first|: given two collections of objects, masks all objects in the first collections that are closer than a predefined value in $\Delta R^2 = \Delta \eta^2 + \Delta \phi^2$ to an object in the second collection
  \item \verb|select_opposite_sign_muons|: given a collection of objects with charge (i.e. muons), masks all but the first two objects ordered by $p_T$ which are of opposite charge
\end{itemize}

\begin{figure}[!tp]
  \centering
  \includegraphics[width=1.0\linewidth]{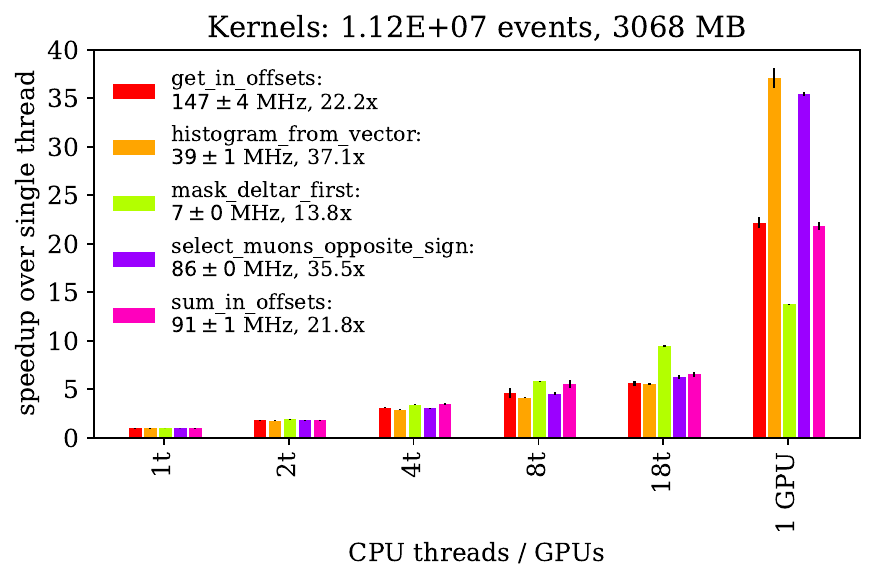}
  \caption{Benchmarks of the computational kernels. We compare the performance of the kernels on approximately 11 million preloaded events with respect to the number of CPU threads used. We find that using multiple CPU threads leads to a sublinear increase in performance, whereas the kernels on the GPU generally receive a ~20-30x speedup over a single thread. In particular, we find that the kernel for computing $\Delta R$ masking between two collections runs at a speed of 7 MHz on a single-thread of the CPU and is sped up by about a factor x14 using the GPU.}
  \label{fig:kernel_benchmarks}
\end{figure}

The kernels have been implemented in Python and just-in-time compiled to either multithreaded CPU or GPU (CUDA) code using the Numba package~\cite{lam2015numba}.
We have chosen Python and Numba to implement the kernels in the spirit of quickly prototyping this idea, but this approach is not restricted to a particular programming environment.
The total number of lines of code for both the CPU and GPU implementations of all kernels is approximately 400, reflecting the relative simplicity of the code.
We have benchmarked the performance of the kernels on preloaded data on a multicore GPU-equipped workstation\footnote{24 core Intel Xeon E5-2687W v4 @ 3.00GHz, an nVidia Geforce Titan X GPU, an Intel Optane 900P for caching and networked storage using CephFS connected over a 10Gbit/s LAN}.
The results are shown on Fig.~\ref{fig:kernel_benchmarks}.
We find that even complex kernels perform at speeds of tens of MHz (millions of events per second). We find a total speedup of approximately 20-30x over single-core performance on a GPU for the kernels.

\section{Analysis benchmark}
\label{sec:performance}
Besides testing the individual kernels in a controlled setting, we benchmark this kernel-based approach in a prototypical top quark pair analysis using CMS Open Data from 2012~\cite{cmsopendata-mc,cmsopendata-data}. The datasets are processed from the experiment-specific format to a columnar format using a publicly-available tool~\cite{aod2nanoaod}. We stress that the output of this tool is not meant to be used for deriving physics results, but rather to replicate the computing conditions that are encountered in the data analysis groups at the experiments. The resulting derived datasets, corresponding to about 60GB of simulation and 40GB of data with the single muon trigger, are further processed in our benchmark analysis.

The benchmark analysis implements the following features in a single end-to-end pass:
\begin{itemize}
    \item event selection based on event variables: trigger bit, missing transverse energy selection
    \item object selection based on cuts on objects: jet, lepton selection based on $p_T$, $\eta$
    \item matching of pairs of objects: jet-lepton $\Delta R$ matching, jet to generator jet matching based on index
    \item event weight computation based on histogram lookups: pileup, lepton efficiency corrections
    \item jet energy corrections based on histogram lookups: \textit{in-situ} jet energy systematics, increasing the computational complexity of the analysis by about 40x
    \item high-level variable reconstruction: top quark candidate from jet triplet with invariant mass closest to $M=172$ GeV
    \item evaluation of $\simeq 40$ typical high-level inputs to a deep neural network (DNN)
    \item Multilayer, feedforward DNN evaluation using \texttt{tensorflow}
    \item saving all DNN inputs and outputs, along with systematic variations, to histograms ($\simeq 1000$ individual histograms)
\end{itemize}
We find that performing systematic corrections based on linear interpolations, re-evaluating the DNN classifier and filling thousands of histograms are the most computationally demanding steps of the analysis.
Generally, this code represents a simplified but prototypical Standard Model analysis. We expect further work on analyses with LHC Open Data to result in more realistic codes. We note that ongoing analyses at CMS have already started to adapt to such array-based methods.

We perform two benchmark scenarios of this analysis: one with a partial set of systematic uncertainties to emulate a simpler IO-limited analysis, and one with the full set of systematic uncertainties to test a compute-bound workload. The timing results from the benchmark are reported in table~\ref{tab:benchmarks}. Generally, we observe that the simpler analysis can be carried out at rate of approximately 50 kHz / CPU-thread, with the GPU-accelerated version performing about 10x faster than a single CPU thread.

On the other hand, a complex compute-bound analysis where around the main workload is repeated around 40x is about 15x faster on a single GPU than on 1 CPU thread.
This modest advantage compared to the 20-30x improvement in raw kernel performance can be attributed to suboptimal kernel scheduling, as we have observed the GPU utilization peak around 50\% in typical benchmarking.
A single pass of the analysis involves hundreds to thousands of kernel calls, such that further optimizations in kernel scheduling and fusion are needed to keep the accelerator fully loaded. We find that it is beneficial to ensure that the kernels are called on sufficiently large datasets to reduce the overhead of kernel scheduling with respect to the time spent in the computation.

Although the approximately 10-15x performance improvement on the GPU with respect to a single CPU thread is relatively modest, it is promising to see that most of the physics analysis methods can be implemented with relative ease on a GPU, such that with further optimizations, a small number multi-GPU machines might be a viable alternative to a large server farm in the future. Ultimately, the choice of the computing platform will be driven by availability and pricing of resources.  We stress that we do not claim a GPU is necessarily faster than a large number of CPU threads, but rather demonstrate that it is possible to implement a portable end-to-end analysis on various backends by using relatively standard tools.

\begin{table}[!t]
\centering
\caption{Processing speed and core-hours to process one billion events for the analysis benchmark with partial and full systematics.}
\label{tab:benchmarks}
\begin{tabular}{c|cc}
job type & partial systematics & full systematics \\
\hline
\hline
\multicolumn{3}{c}{processing speed (kHz)} \\
\hline
1 thread & 50 & 1.4 \\
4 threads & 119 & 4.0 \\
GPU & 440 & 20 \\
\hline
\multicolumn{3}{c}{walltime to process a billion events (hours)} \\
\hline
1 thread & 5.5 & 200 \\
4 threads & 2.3 & 70 \\
GPU & 0.6 & 13 \\
\hline
\end{tabular}
\end{table}

% \begin{figure}[!tp]
%   %\includegraphics[width=1.0\linewidth]{jagged.pdf}
%   \centering
%   \includegraphics[width=0.8\linewidth]{analysis_benchmark.pdf}
%   \caption{Benchmarks of the full analysis with accelerated kernels on 1 thread, 4 threads as well as a GPU, comparing the processing speed of the analysis with partial systematic uncertainties (blue) the one with 40x systematic variations (orange). In the legend, we show the absolute single-thread performance along with the GPU speedup.}
%   \label{fig:analysis_benchmark}
% \end{figure}

We have not performed a symmetrical benchmark with respect to current commonly used ROOT-based analysis codes, as these cannot be easily run from a single integrated workflow and thus cannot be reliably benchmarked end-to-end. However, neglecting the time spent in batch job preparation and management, event processing speeds for ROOT-based event loop analyses are commonly in the range of 10-50 kHz / CPU thread, which also factors in data fetching and decompression. Neither multithreading nor GPU offloading are generally possible in these codes. We therefore stress that the benchmark shown in this result is not aimed to be symmetrical against standard ROOT-based analysis codes, but rather reflects the possibilities afforded by using array-based techniques on local data for analysis, which allows for fast experimentation and iteration on sizeable datasets before resorting to extensive distributed computing infrastructure.

\section{Summary and outlook}
\label{sec:summary}
We demonstrate that it is possible to carry out prototypical end-to-end high-energy physics data analysis from a relatively simple code on a using multithreading to process millions of events per second on a multicore workstation.
This is achieved with memory-mappable caching, array processing approaches and a small number of specialized kernels for jagged arrays implemented in Python using Numba. It is also possible to offload parts of these array computations to accelerators such as GPUs, which are highly efficient at parallel processing and alleviate compute-bound analyses such as those with many systematic variations or DNN evaluations. We demonstrate a prototypical top quark pair analysis implementation using these computational kernels with optional GPU offloading that can run between 20 - 400 kHz on a single GPU, about 10x faster than a single CPU thread.

Several improvements are possible, among them optimizing data access via efficient IO, using accelerators for data decompression, scaling horizontally across multiple machines using scheduling software such as Apache Spark or dask, scaling vertically in a single server by optimizing the threading performance as well as kernel scheduling and fusion on heterogeneous architectures. Setting up integrated "analysis facilities" at HPC centers, with relevant datasets available on a high-performance filesystem coupled with multithreaded processors and GPUs would allow high-energy physics experiments to iterate through the late-stage data without the typical overheads from distributed computing. We hope that this report sparks further discussion and development of efficient and easy-to-use analysis tools which would be useful for scientists involved in HEP data analysis and in other data intensive fields.

\section*{Acknowledgments}
The excellent performance of the LHC as well as the CERN computing infrastructure and support were essential for this project. The authors are grateful to the CMS collaboration for providing the 2012 dataset as Open Data, especially the Higgs and Computing groups and the USCMS computing group at Fermilab, as well as Guenther Dissertori for useful discussions. We would like to thank Jim Pivarski and Lindsey Gray for helpful feedback at the start of this project and Christina Reissel for being an independent early tester of these approaches.
The availability of the excellent open-source Python libraries \texttt{uproot}, \texttt{awkward}, \texttt{coffea}, \texttt{numba}, \texttt{cupy} and \texttt{numpy} was imperative for this project and we are grateful to the developers and maintainers of these open-source projects.
This work was partly conducted at "\textit{iBanks}", the AI GPU cluster at Caltech.
We acknowledge NVIDIA, SuperMicro  and the Kavli Foundation for their support of "\textit{iBanks}".
This work is supported by the the Prime NSF Tier2 award 1624356 and the Caltech HEP DOE award DE-SC0011925.

\end{document}